\documentclass[12pt]{article}

\usepackage{amsmath,amsfonts,amsthm,amssymb}
\usepackage{epsfig}  
\usepackage{feynmf}
\def\ii{\'\i}

\def\cao{\c c\~ao}

\def\ii{\'\i}

\def\cao{\c c\~ao}

\def\ftoday{{\sl {Le \number\day \space\ifcase\month 
\or janvier\or f\'evrier\or mars\or avril\or mai
\or juin\or juillet\or ao\^ut\or septembre\or octobre
\or novembre \or d\'ecembre\fi\space \number\year}}}    
\def\ptoday{{\sl {\number\day \space de\space \ifcase\month 
\or janeiro\or fevereiro\or mar{\c c}o\or abril\or maio
\or junho\or julho\or agosto\or setembro\or outubro
\or novembro \or dezembro\fi\space de\space \number\year}}}    
\def\gtoday{{\sl {Den \number\day. \ifcase\month 
\or Januar\or Februar\or M\"arz\or April\or Mai
\or Juni\or Juli\or August\or September\or Oktober
\or November \or Dezember\fi\space \number\year}}}    
\def\today{{\sl {\ifcase\month
\or January\or February\or March\or April\or May
\or June\or July\or August\or September\or October
\or November \or December\fi \space\number\day,\space 
                                            \number\year}}}
\newcommand{\journal}[4]{{\em #1~}#2\,(#3)\,#4}

\newcommand{\jhep}{\journal {J. High Energy Phys.}}

\newcommand{\pr}{\journal {Phys. Rev.}}

\newcommand{\prl}{\journal {Phys. Rev. Lett.}}

\newcommand{\jpcs}{\journal {J. Phys. Conf. Ser.}}

\newcommand{\cqg}{\journal {Class. Quantum Grav.}}

\newcommand{\np}{\journal {Nucl. Phys.}}
\newcommand{\pl}{\journal {Phys. Lett.}}

\newcommand{\jgp}{\journal {J. Geom. Phys.}}

\newcommand{\epj}{\journal {Eur. Phys .J.}}

\renewcommand{\a}{\alpha}
\renewcommand{\b}{\beta}
\newcommand{\g}{\gamma}           
\renewcommand{\d}{\delta}         
\newcommand{\e}{\varepsilon}

\newcommand{\la}{\lambda}        \newcommand{\LA}{\Lambda}
\newcommand{\m}{\mu}
\newcommand{\n}{\nu}
\newcommand{\om}{\omega}         
\newcommand{\p}{\psi}              

\newcommand{\f}{{\phi}}           

\newcommand{\CC}{{\cal C}}

\newcommand{\MM}{{\cal M}}

\newcommand{\SSS}{{\cal S}}

\newcommand{\TT}{{\cal T}}

\newcommand{\es}{\\[3mm]}

\newcommand{\sla}{\raise.15ex\hbox{$/$}\kern -.57em} 
\newcommand{\Sla}{\raise.15ex\hbox{$/$}\kern -.70em}

\def\Lp{\displaystyle{\biggl(}}
\def\Rp{\displaystyle{\biggr)}}

\newcommand{\lp}{\left(}\newcommand{\rp}{\right)}

\newcommand{\complex}{{\kern .1em {\raise .47ex
\hbox {$\scriptscriptstyle |$}}
    \kern -.4em {\rm C}}}
\newcommand{\real}{{{\rm I} \kern -.19em {\rm R}}}
\newcommand{\rational}{{\kern .1em {\raise .47ex
\hbox{$\scripscriptstyle |$}}
    \kern -.35em {\rm Q}}}
\renewcommand{\natural}{{\vrule height 1.6ex width
.05em depth 0ex \kern -.35em {\rm N}}}

\newcommand{\half}{\frac{1}{2}}
\newcommand{\pa}{\partial}

\newcommand{\dfud}[2]{{\displaystyle{\frac{\delta #1}{\delta #2}}}}

\newcommand{\dint}{\displaystyle{\int}}
\newcommand{\eg}{{\em e.g.,\ }}

\newcommand{\ie}{{{\em i.e.},\ }}

\newcommand{\twiddle}{\lower.9ex\rlap{$\kern -.1em\scriptstyle\sim$}}

\newcommand{\equ}[1]{(\ref{#1})}
\newcommand{\eq}{\begin{equation}}
\newcommand{\eqn}[1]{\label{#1}\end{equation}}
\newcommand{\eea}{\end{eqnarray}}
\newcommand{\eqa}{\begin{eqnarray}}
\newcommand{\eqan}[1]{\label{#1}\end{eqnarray}}
\newcommand{\ba}{\begin{array}}
\newcommand{\ea}{\end{array}}
\newcommand{\eqac}{\begin{equation}\begin{array}{rcl}}
\newcommand{\eqacn}[1]{\end{array}\label{#1}\end{equation}}

\makeatletter 
\@addtoreset{equation}{section}  
\makeatother 
 
\newcommand{\bz}{\begin{enumerate}}
\newcommand{\ez}{\end{enumerate}}

\newcommand{\ADS}{(A)dS}

\begin{document}

\title{Chern-Simons Gravity in Four Dimensions} 

\author{Ivan Morales, Bruno Neves, 
Zui Oporto\footnote{Present address: Carrera de F\ii sica, 
Universidad Mayor de 
San Andr\'es, La Paz, Bolivia.} and 
Olivier Piguet
\\[4mm]
{\small Departamento de F\ii sica, Universidade Federal de 
Vi\c cosa (UFV)}\\
{\small  Vi\c cosa, MG, Brazil}
}
\date{January 2017}    

\maketitle

\vspace{-5mm}

\begin{center}
{\small\tt E-mails:
mblivan@gmail.com, bruno.lqg@gmail.com, \\
azurnasirpal@gmail.com, opiguet@pq.cnpq.br }
\end{center}

\vspace{3mm}

\begin{abstract}
Five dimensional Chern-Simons theory with 
(anti-)de Sitter SO(1,5) or SO(2,4)
gauge invariance presents an alternative to General Relativity
with cosmological constant.
We consider the zero-modes of its Kaluza-Klein compactification 
to four dimensions. 
Solutions with vanishing torsion are obtained 
in the cases of a spherically symmetric 3-space
and of a homogeneous and isotropic 3-space, which reproduce the 
Schwarzshild-de Sitter and
$\Lambda$CDM cosmological solutions of General Relativity.
We also check that vanishing torsion is a stable feature of the solutions.

\end{abstract}

Keywords: Topological gravity; General Relativity; Cosmology;
 Higher di\-men\-sions.

PACS numbers: 04.20.Cv, 04.50.Cd


\section{Introduction}

Our present understanding of the fundamental processes in Nature
is dominated by two extremely efficient theories: the already 
half a century old Standard Model (SM) valid in the quantum
microscopic realm, and the centenary General Relativity (GR) 
valid in the classical macroscopic realm, from GPS 
monitoring in the planetary scale up to the cosmic 
scale, the evolution of the Universe from the Big Bang 
up to an unforeseeable future. No observation neither 
experiment have shown any falsification of both theories, 
up to now\footnote{See however~\cite{LHCb} for an experimental result hinting to a possible problem with the Standard Model.}.

 An important problem, however, is theoretical: the contradiction of
GR being classical and SM being quantum. Two cures may be conceived.
The more radical one may be the construction of a new framework, 
beyond ``quantum'' and ``classical'', in which 
GR and SM would stay as approximations of a unique 
theory, each being valid in its respective domain. String theory represents an effort in this direction. 

Another, more obvious and (only apparently) straightforward 
cure is the direct quantization of GR, along the 
canonical lines of Loop Quantum 
Gravity~\cite{Rovelli-Thiemann}, for instance.
The latter is based on a first order formulation of GR, 
which has two local symmetries: the invariances under 
the space-time diffeomorphisms and the local Lorentz transformations. 
In  Dirac's canonical formalism~\cite{Dirac}, 
a constraint  is associated to each local invariance, 
which has to be solved at the quantum level. Unfortunately, 
one of  these constraints, namely the one associated 
 with the time 
diffeormophism invariance -- called the 
hamiltonian or scalar constraint -- has resisted to any tentative 
of solving it, up to now -- although important 
progresses have been 
made~\cite{WdW,scalar-constr,Rovelli-Vidotto}.

It happens that the de Sitter or anti-de Sitter (\ADS)
gravitation theory in $5D$ space-time 
defined by
a Chern-Simons theory with the 
gauge groups SO($n$,$6-n$) for $n$ = 1 or 2~\cite{hassaine-zanelli}
 shows the remarkable 
property of its time-diffeomorphism constraint being a 
consequence of its gauge invariance and its invariance under the
space diffeomorphisms~\cite{Banados-etal}. 
It follows that the scalar constraint is then 
an automatic consequence
of the other ones. This yields a first motivation for studying
this particular theory of gravity.

A second motivation is given by the fact that the presence of a cosmological constant, hence of a fundamental scale
at the  classical level, happens as a necessary feature 
of this theory, as we shall verify, in contrast with usual GR
where its presence or not is the result of an arbitrary choice.

The Chern-Simons \ADS\ theory is a special case of the 
extensions of Einstein theory known as Lovelock 
theories~\cite{Lovelock} which, despite of containing
higher powers of the curvature, obey second order field equations.
There exists a vast literature\footnote{Only a few references are given here. A rather complete list may be found
in the book~\cite{hassaine-zanelli}, which offers an up-to-date
 review on Lovelock and 
Chern-Simons theories of gravitation.} 
on Lovelock theories, beginning with 
the historical papers~\cite{Boulware-Deser,Zumino,Wheeler}.
The paper~\cite{Dianyan} already gives explicit solutions of the 
Schwarzschild, Reissner-Nordstr\"om and Kerr type in higher dimension
Einstein theory with cosmological constant. More recent works may
be divided into general Lovelock 
models~\cite{Banados2004,Cai-Ohta,Canfora2007,Canfora2016,%
Castillo-etal},
Chern-Simons model based on \ADS\ gauge invariance~\cite{Aros2008}, 
and Chern-Simons models based on larger gauge 
groups, see in particular~\cite{Salgado1}--\cite{Salgado4}. 
It is worth noticing the work of\cite{Salgado3}, where the choice of
the gauge group extension leads to a theory which reduces to 5D
Einstein  theory with cosmological constant in 
the case of a vanishing torsion. We may also mention 
genuinely 4-dimensional models together with
the search for physically reliable solutions of
them, such as the Chamseddine 
model~\cite{Chamseddine,topgrav-MNOP} 
obtained from the 5D \ADS\ Chern-Simons
by dimensional reduction and truncation of some fields, 
or the model of~\cite{Toloza-Zanelli} obtained by adding 
to the Einstein-Hilbert action the coupling of a scalar field with
the 4D Euler density. 

The aim of the present work is an  investigation of the 
classical properties of the 4D theory obtained from the 5D \ADS\ 
Chern-Simons theory 
by a Kaluza-Klein compactification, find solutions of the field equations
with spherical symmetry and solutions of the cosmological type,
and comparisons of these solutions with the results
 of usual GR. This is
intended to be a preliminary 
step to any attempt of quantization, the latter deserving future care.

\ADS\ theory and its reduction to 4 dimensions are reviewed in 
Section \ref{ADS-review}, solutions with spherical symmetry 
and cosmological solutions are showed in 
Sections \ref{spherical-sol} 
and \ref{cosmology-sol}. Conclusions are presented in 
Section \ref{conclusion}.  Appendices present details omitted in the main text.

\section{(A)dS Chern-Simons theory for 5D and 4D \\
gravity}\label{ADS-review}
\subsection{\ADS\ Chern-Simons theory as a 5D gravitation theory}

Apart of some considerations from the authors, the content of this subsection is not new.
A good review may be found in
the book~\cite{hassaine-zanelli} together with references to the original literature\footnote{ Notations and conventions are given in 
Appendix \ref{not-conv}}. 

Chern-Simons theories are defined in odd-dimensional space-times, 
we shall concentrate to the 5-dimensional case. We first define the gauge group as  
the pseudo-orthogonal group SO(1,5) or SO(2,4), the de 
Sitter or anti-de Sitter group in 5 dimensions, generically denoted by \ADS. These are the matrix groups leaving invariant the quadratic forms 
\[\ba{l}
\eta_{\rm\ADS} = \mbox{diag\ }(-1,1,1,1,s)\,,\es s=1 \mbox{ for de Sitter}\,,\quad
  s=-1 \mbox{ for anti-de Sitter}\,.
\ea\]
A convenient basis of the Lie algebra of \ADS\ is given by 10 Lorentz 
SO(1,4) generators $M_{AB}$ = $-M_{BA}$ and 5 ``translation'' generators
$P_A$, where  $A$, $B$, etc., are Lorentz indices taking the values 
$0,\,\cdots,\,4$. These generators obey the commutation rules
\eq\ba{l}
[M_{AB},M_{CD}]=  \eta_{BD} M_{AC} + \eta_{AC} M_{BD}
- \eta_{AD} M_{BC} - \eta_{BC} M_{AD}\,,\\[2mm]
[M_{AB},P_C] = \eta_{AC} P_B - \eta_{BC} P_A\,,
\quad { [P_A,P_B] =- s\,M_{AB}}\,,
\ea\eqn{AdS-algebra}
 where $\eta_{AB}:=\mbox{diag}(-1,1,1,1,1)$   
is the $D=5$ Minkowski metric.

We then define the \ADS\ connexion 1-form, expanded in this basis 
as
\eq
{A}(x) = \frac12 {\om}^{AB}(x) M_{AB} + \dfrac{1}{ l}\, {{e}}^A(x) P_A\,,
\eqn{expand-connection}
where ${l}$ is a parameter of dimension of a length. 
$\om^{AB}$ will play the role of the 5D Lorentz connection form  and ${{e}}^A$ of the 
``5-bein'' form in the corresponding gravitation theory. 
We may already note that the presence of the parameter $l$, which will be related 
to the cosmological constant (see Eq. \equ{Lambda-G5}), is necessary in order to match the dimension of 
the 5-bein form ${{e}}^A$, which is that of a length, 
to that of the dimensionless Lorentz connection form
${\om}^{AB}$.

The \ADS\ gauge transformations of the connection read, in infinitesimal
form, as
 \[
\d {A} = d\epsilon -  [{A},\epsilon]\,,
\]
where the infinitesimal 
parameter $\epsilon$ expands as
\[
\epsilon(x) = \frac12 {\epsilon}^{AB}(x) M_{AB} 
+ \dfrac{1}{ l}\, {\b}^A(x) P_A\,.
\]
From this follow the transformations rules of the fields $\om$
and ${e}$:
\eq\ba{l}
\d{\om}^{AB} = d\epsilon^{AB} +{\om}^A{}_ C \,\epsilon^{CB}
+{\om}^B{}_C \,\epsilon^{AC}
-\dfrac{s}{l^2}\lp {e}^A \b^B - {e}^B \b^A   \rp\,,
\es
\d{e}^A = {e}_C\,\epsilon^C{}^A + d\b^A +{\om}^A{}_C\,\b^C \,.
\ea\eqn{gauge-transf}

Desiring to construct a background independent theory, we assume a dimension 5 manifold $\MM_5$ without an a priori metric. Then the unique \ADS\  gauge invariant
action -- up to boundary terms --  which may 
constructed with the given connection is the Chern-Simons action for the group \ADS,
which in our notation reads
\eq\ba{l}
S_{\rm CS} = \dfrac{1}{8\kappa}\dint_{\!\!\!\!\MM_5} \e_{ABCDE} \Lp 
{{e}}^A\wedge {{R}}^{BC}\wedge {{R}}^{DE}
-\dfrac{2s}{3 l^2} {{e}}^A\wedge {{e}}^B\wedge {{e}}^C\wedge {{R}}^{DE}\\[2mm]
\phantom{S = \frac{\kappa'}{8}\int \e_{ABCDE} \Lp}
+ \dfrac{1}{5 l^4} {{e}}^A\wedge {{e}}^B\wedge {{e}}^C\wedge {{e}}^D\wedge {{e}}^E
\Rp\,,
\ea\eqn{CS-grav-action}
where  $\kappa$ is a dimensionless\footnote{In our units 
$c=1$.} coupling constant and
\eq 
{{R}}^A{}_B=d{{\om}}^A{}_B+{{\om}}^A{}_C\wedge{{\om}}^C{}_B
\eqn{Riemann}
 is the Riemann curvature 2-form associated to the Lorentz connection 
${\om}$.
We may add to the action  a part $S_{\rm matter}$ describing matter and its interactions
with the geometric fields ${{\om}}^{AB}$ and ${{e}}^A$, which leads to a total action
$S=S_{\rm CS} $ + $S_{\rm matter}$. 
The resulting  field equations read
\eq\ba{l}
\dfud{S}{{{e}}^A} = \dfrac{1}{8\kappa} \e_{ABCDE}{{F}}^{BC}\wedge {{F}}^{DE}
+ {\TT}_A = 0 \,,\es
\dfud{S}{{{\om}}^{AB}}=
  \dfrac{1}{2\kappa} \e_{ABCDE}{{T}}^C\wedge {{F}}^{DE}
+ {\SSS}_{AB}  =0 \,,
\ea\eqn{field-eq}
where 
\eq
{{T}}^A = D {{e}}^A =
 d {{e}}^A +{{\om}}^A{}_B\wedge {{e}}^B
\eqn{torsion}
 is the torsion 2-form, 
\eq
{{F}}^{AB}={{R}}^{AB} -\dfrac{s}{l^2} {{e}}^A\wedge {{e}}^B
\eqn{FAB}
 is the \ADS\ curvature, and
\eq
{\TT}_A :=  \dfud{ S_{\rm matter}}{ {{e}}^A}\,,\quad
{\SSS}_{AB} := \dfud{S_{\rm matter}}{{{\om}}^{AB}}
\eqn{matter}
are  the energy-momentum 4-form, related to the energy-momentum components ${\TT}^A{}_B$ in 
the 5-bein frame, by
\eq
{\TT}_A =\frac{1}{4!}\e_{BCDEF} {\TT}^B{}_A \,
{{e}}^C \wedge {{e}}^D \wedge {{e}}^E \wedge {{e}}^F\,,
\eqn{em-4-form}
and the spin 4-form ${\SSS}_{AB}$.

A generalized continuity equation for energy, momentum and 
spin results from the field equations \equ{field-eq}, the
zero-torsion condition $D{e}{}^A=0$ and the Bianchi identity
$D{{R}}^{AB}=0$:
\eq
D{\TT}_A + \frac{s}{l^2}\, {\SSS}_{AB} \wedge {{e}}^B = 0\,,
\eqn{cont-eq}
which reduces to the energy-momentum continuity equation
in the case of spinless matter:
\eq
D{\TT}_A = 0\,.
\eqn{spinless-cont-eq}

We observe that the sum of the second and third term of the 
action \equ{CS-grav-action} 
 is proportional to the 5D Einstein-Palatini action with 
 cosmological constant, 
 which is  equivalent to 
 the more familiar  5D  Einstein-Hilbert action in the metric formulation:
 \[
S_{\rm EH} = \dfrac{1}{16\pi G_{(5D)}}\,
\dint_{\!\!\!\!\MM_5} d^5x \sqrt{- g} 
({R}-2\LA)\,,
 \]
 with $G_{(5D)}$ the 5D gravitation constant, $\LA$ 
 the cosmological constant,  ${R}$ the Ricci scalar and 
 $ g$ = $-\mbox{det}({{e}}^A{}_\a)^2$ the determinant 
 of the 5D metric
 \eq
 { g}_{\a\b } = \eta_{AB} {{e}}^A{}_\a {{e}}^B{}_\b\,.
 \eqn{5D-metric}
This allows us to express the parameters $\kappa$ and $l$ in terms of the 
5D physical parameters $G_{(5D)}$ and $\LA$ as
\eq
\dfrac{3s}{l^2}=\LA \,, \quad
 \kappa = -\dfrac{8\pi}{3} \LA \,G_{(5D)}\,.
\eqn{Lambda-G5}
The coefficient of the first term in the 
action \equ{CS-grav-action} -- the so-called Gauss-Bonnet 
term -- is of course fixed  by \ADS\ 
gauge invariance in terms of the two parameters of the theory. 
This is a special case of the more general Lanczos-Lovelock or
Lovelock-Cartan theory~\cite{lovelock-cartan,hassaine-zanelli}

\subsubsection{A trivial solution}

In the vacuum defined by the absence of matter,
a special class of solutions of the field equations 
\equ{field-eq} is that of the solutions of the 
 stronger equations
\[
F^{AB} = 0\,,
\]
with the \ADS\ curvature 2-forms $F^{AB}$ given by \equ{FAB}.
In fact the solution is unique  up to an arbitrary torsion
as it is readily seen 
by inspection of the second of the field equations 
\equ{field-eq}. This is a solution of constant curvature
and corresponds to an empty de Sitter or anti-de Sitter 
5D space-time with a 5-bein form
\eq\ba{l}
{{e}}^A =  \sqrt{1-\frac{\LA}{3} r^2} \, dt 
+  \dfrac{1}{\sqrt{1-\frac{\LA}{3} r^2}}\, dr 
+ r\lp d\theta + \sin\theta \, d\f 
      + \sin\theta\sin\f\, d\p \rp\,,
\ea\eqn{5-bein-ads5}
or its Lorentz transforms, leading to the metric
\eq\ba{l}
ds^2 = - \lp 1-\frac{\LA}{3} r^2\rp dt^2 
+  \dfrac{dr^2}{1-\frac{\LA}{3} r^2} 
+ r^2\lp d\theta^2 + \sin^2\theta \, d\f^2 
      + \sin^2\theta\sin^2\f\, d\p^2 \rp\,,
\ea\eqn{metric-ads5}
in spherical 4-space coordinates $t,\,r,\,\theta,\,\f,\,\p$.
This metric has the symmetry O(4) of 4-space rotations.

\subsection{Compactification to 4 dimensions}

In order to connect the theory with 4-dimensional physics, we choose to implement
a Kaluza-Klein type of compactification~\cite{Overduin-Wesson}, considering the 4th
spatial dimension to be compact. In other words, we consider a 5D space-time with the
topology of $\MM_5$ = $\MM_4$ $\times$ $S^1$,  
the first factor being a 4-dimensional manifold
and $S^1$ the circle representing the compactified dimension\footnote{The 
coordinates of $\MM_5$ are denoted by 
$x^\a$ ($\a=0,\cdots,4$) and those of $\MM_4$  by 
$x^\m$ ($\m=0,\cdots,3$). The coordinate of $S^1$
is denoted by $\chi$, with $0\leq\chi< 2\pi$.}. 
Any space-time function admits 
a Fourier expansion in the $S^1$ coordinate $\chi$,  its coefficients --
the Kaluza-Klein modes -- being functions in $\MM_4$.  
In the applications presented in Sections \ref{spherical-sol} 
and \ref{cosmology-sol} only the zero mode is considered, which amounts 
to consider all functions as constant in $\chi$.

 Note that the zero \ADS\ curvature solution 
(\ref{5-bein-ads5},\ref{metric-ads5}) is not a solution 
of the compactified theory.

\subsection{Solutions with zero torsion}
\subsubsection{On the number of degrees of freedom}\label{number-dof}

The number of local physical degrees of freedom of the theory is best 
calculated by means of a canonical analysis.
It is 
known~\cite{Banados-etal}
that the present theory has 75 constraints of first and second class.
The number $n_2$ of the latter is equal, in the weak 
sense\footnote{A 
``weak equality'' is an equality valid up to the
constraints.},  to the rank $r$ of the matrix
formed by the Poisson brackets of the constraints. The number of first class constraints -- which generate the gauge transformations -- is thus equal to $n_1$ = $75-n_2$. Moreover, the number of generalized 
coordinates\footnote{The generalized coordinates are the space components
of the connection and 5-bein fields: ${\om}^{AB}{}_a$ and ${e}^A{}_a$, 
with $a=1,\,\cdots,\,4$.}
 is equal to 60. Thus the number $n_{\rm d.o.f.}$ of 
 physical degrees of freedom, at each point of 4-space, 
 is given by
\[
n_{\rm d.o.f.} = \half\lp 2\times60 -2n_1-n_2 \rp = \dfrac{r}{2} -15\,.
\]
The authors of~\cite{Banados-etal} have shown that the result for the rank
$r$, hence for $n_{\rm d.o.f.}$, depends on the region of phase space
where the state of the system lies. They have computed it 
in the ``generic'' case, \ie the case where the rank $r$ 
is maximal, corresponding to the situation with the minimal set of local
invariances, namely that of the fifteen \ADS\ gauge invariances and the
four  4-space diffeomorphism invariances. This results in $r=56$, \ie
in 13 physical degrees of freedom.

The case with zero torsion is non-generic in the sense given above.
We have checked by numerical tests that the rank $r$ is then at most 
equal to 40, which shoes that $n_{\rm d.o.f.}\leq5$  
in the case of a zero torsion.

\subsubsection{On the stability of solutions with zero 
torsion}\label{stability-criterion}

The second of the field equations \equ{field-eq} is identically solved 
by assuming zero torsion. We would like to know in which extent 
solutions with zero torsion are stable under small perturbations.
More precisely,  considering a field configuration with a 
torsion of order $\epsilon$, we will look 
for conditions ensuring its vanishing as a 
consequence of the equations. 

The second of Eqs. \equ{field-eq}, written in 5-bein components as
\[
\dfrac{1}{8} \e_{ABCDE}\,\e^{XYZTU}\,{T}^C{}_{XY}\,{F}^{DE}{}_{ZT}
 = 0\,,
\]
can be rewritten in the form
\eq
{T}^i\, M_i{}^j  = 0\,,
\eqn{eq-torsion}
with 
\eq
M_i{}^j = 
\dfrac{1}{8} \e_{ABCDE}\,\e^{XYZTU}\,{F}^{DE}{}_{ZT}\,, 
\eqn{def-M}
the index
 $i$ standing for $(C,\,[XY])$ and $j$ for $(U,\,[AB])$. If the $50\times50$
 matrix $M$ is inversible, then  \equ{eq-torsion} 
 implies the vanishing of the torsion. 
 
Let us write the infinitesimal torsion as ${T}^A{}_{BC}$ = 
$\epsilon\,t^A{}_{BC}$. We note that the connection 
${\om}$  \equ{full-connection}
constructed from the 5-bein and the torsion is 
linear in the torsion components, thus in $\epsilon$, hence
${F}$ is a polynomial in $\epsilon$, and so is the matrix $M$. 
This implies
that its inverse $M^{-1}$ exists and is analytic in $\epsilon$ in a 
neighborhood of $\epsilon=0$, if the matrix
\eq
M_i{}^{j(0)} = \left. M_i^j\right|_{\epsilon=0}
\eqn{matrix-M0}
is regular. It then follows, under the latter assumption,
that the torsion vanishes.
We can summarize this result as the following

\noindent{\bf Stability criterion:} A sufficient condition for the stability
of the solutions at zero torsion under possible fluctuations 
of the torsion
is that the matrix \equ{def-M} restricted to
zero torsion, $M^{(0)}$, be regular.
  
 This criterion is important in view of the difference between the number of physical degrees of freedom for states with zero torsion and this number for generic states\footnote{We thank Jorge Zanelli for pointing out 
this problem to us.}, as discussed in 
Subsection \ref{number-dof}. 
Indeed, if the state of the system lies in the sub-phase space of zero torsion states, the fulfillment of the condition of the criterion guaranties  that the state will evolve staying in that subspace.

%
%

\section{Solutions with 3D rotational symmetry}\label{spherical-sol}

The most general metric and torsion tensor components 
compatible with the rotational symmetry of 3-space are 
calculated in 
Appendix \ref{app-spher}, with the metric given by 
\equ{metric-t-r-chi} and the torsion by \equ{ss} in 
a system of coordinates $t,\,r,\,\theta,\,\phi,\,\chi$. 
All component fields depend on $t,\,r,\,\chi$. But 
we shall restrict ourselves here to look
for stationary solutions, neglecting also the higher 
Kaluza-Klein modes. Thus only a dependence on the 
radial coordinate $r$ is left. In this situation the metric
takes the simpler form \equ{mss1} with only one non-diagonal term, thanks to some suitable coordinate transformations, as explained 
in Appendix \ref{app-spher}.

Through the definition \equ{5D-metric}, this metric leads 
to the 
5-bein $e^A$ = $e^A{}_\a dx^\a$, up to local Lorentz transformations 
$e'{}^A=\LA^A{}_B e^B$, with
\eq 
(e^A{}_\a) =  \lp\ba{ccccc}
n(r)&0&0&0&c(r)\\   
0&a(r)&0&0&0)\\  
0&0 &r &0&0\\ 
0&0&0&r\sin\theta&0\\
0&0&0&0&b(r)
\ea\rp \,,
\eqn{5bein-esferico}
 and the relations
\[
g_{tt}(r)=-n^2(r)\,,\ \ \ g_{rr}(r)=a^2(r)\,,
\ \   g_{t\chi}(r)=-n(r) c(r)\,,\ \ 
g_{\chi\chi}(r)=b^2(r) - c^2(r)\,.
\]
Beyond the spherical symmetry of 3-space, the stationarity 
and the restriction to the zero KK mode, we still make 
the following hypotheses:
\begin{enumerate}
\item[(i)] The torsion \equ{torsion} is zero: ${T}{}^A=0$,
hence  the second of the field equations \equ{field-eq} is
trivially satisfied;
\item[(ii)]  We look for  static solutions, hence
$g_{t\chi}(r)=0$,  and $c(r)=0$ in \equ{5bein-esferico}. 
\item[(iii)] We restrict the discussion to the de Sitter case,
\ie with a positive cosmological constant: $s=1$, which 
corresponds to the present data~\cite{Planck-data}.
\end{enumerate}
Consistently with the symmetry requirements and the hypotheses above,
the tensor
$\TT^A{}_B$ appearing in the definition \equ{em-4-form} of the  energy-momentum 4-form reads\footnote{ The hats on $\hat\rho$,  etc., mean energy density, etc. in 4-space.}
\eq
\TT^A{}_B=\mbox{diag}\,(-\hat{\rho}(r),\hat{p}(r),
\hat{p}(r),\hat{p}(r),\hat{\la}(r))\,,
\eqn{e-m-tensor}
We also assume that the spin current 4-form ${\SSS}_{AB}$ in \equ{matter}
is vanishing. In the present setting, the continuity equation 
\equ{cont-eq} takes then the form
\eq
p'(r) + p(r)\left(\frac{n'(r)}{n(r)}-\frac{b'(r)}{b(r)}\right)
   -\lambda (r)\frac{ b'(r)}{b(r)}
   +\rho (r)\frac{ b'(r)}{b(r)}=0\,.
\eqn{cont-eq-spherical}
We shall consider the case of an empty physical 3-space, which means
zero energy density and pressure, \ie $\rho(r)=p(r)=0$, keeping only the
``compact dimension pressure''  $\la(r)\not=0$ 
(We shall see that the solution of interest indeed 
has a non-vanishing $\la$).  The continuity equation
thus implies the 5-bein component $b(r)$ to be a constant:
\eq
b(r)=R= \mbox{constant}\,.
\eqn{b=R}
The parameter $R$, which has the dimension of a length, defines the compactification scale.

With all of this, the field equations (first of \equ{field-eq}) reduce
to the three independent equations
\eq\ba{l}
\lp 1-\dfrac{3 r^2}{l^2}\rp a(r)^2 n(r)-2 r n'(r)-n(r)=0 \,,\es
-2 r a'(r)+a(r)^3 \left(\dfrac{3 r^2}{l^2}-1\right)+a(r) =0 \,,\\[4mm]
\kappa  l^2 r^2 a(r)^5 n(r) \lambda(r)
-\lp l^2 a(r) + (r^2-l^2)a(r)^3  \rp n''(r)\es \qquad
+\lp 3 l^2+(r^2-l^2) a(r)^2   \rp a'(r) n'(r)
+2 r a(r)^2 n(r)a'(r)\es \qquad
 -2 r a(r)^3 n'(r) 
  - \lp a(r)^3 +\lp \dfrac{3 r^2}{l^2}-1\rp a(r)^5\rp n(r) = 0 \,.
\ea\eqn{eq-spherical}
Note that these equations do not depend on the 
compactification scale $R$.
The second equation solves for $a(r)$, and then the first one 
yields $n(r)$: 
\eq
n(r)=\sqrt{1-\frac{2\m}{r}-\frac{r^2}{l^2}}\,,
\quad \quad a(r)= 1/n(r)\,,
\eqn{sol-spherical}
after a time coordinate re-scaling is made.  The Schwarzschild mass $\mu$ is an integration constant as in GR.
The third equation yields $\la(r)$ in terms of the functions 
$a(r)$ and $n(r)$, with the final result:
\eq
\la(r) = \frac{6 \mu ^2}{\kappa}\frac{1}{r^6}.
\eqn{lambda-spherical}
The final 5-bein and metric thus read
\eq\ba{l} 
(e^A{}_\a) =  \lp\ba{ccccc}
\sqrt{1-\frac{2\m}{r}-\frac{r^2}{l^2}}&0&0&0&0\\   
0&\lp\sqrt{1-\frac{2\m}{r}-\frac{r^2}{l^2}}\rp^{-1}&0&0&0\\  
0&0 &r &0&0\\ 
0&0&0&r\sin\theta&0\\
0&0&0&0&R 
\ea\rp \,,\\[15mm]
ds^{2} = -(1-\frac{2\m}{r}-\frac{r^2}{l^2}) dt^{2} 
+ \dfrac{dr^2}{1-\frac{2\m}{r}-\frac{r^2}{l^2}} 
+ r^{2}( d\theta^{2} + \sin^{2}\theta\, d\phi^{2})
+R^2 d\chi^{2}  \,.
\ea\eqn{solution-Schw-deS}
This result is just the generalization of 
the Schwarzschild solution in a space-time which is asymptotically 
de Sitter, with cosmologial constant
$\LA$ = $3/l^2$. One remembers that we have described the 
``vacuum'' as described by an energy-momentum tensor 
\equ{e-m-tensor} with one possibly non-zero component: 
the ``compact dimension pressure'' $\la(r)$. Our result 
is that this ``pressure'' is indeed non-vanishing, 
singular at the origin and decaying as the inverse of 
the sixth power of the radial coordinate as shown in 
Eq. \equ{lambda-spherical}.

 We must emphasize that this result follows uniquely from the 
hypotheses we have made.

Finally, we have checked the condition of stability of the 
zero-torsion solutions of the model according to the 
criterion proved in Subsection \ref{stability-criterion}: 
a computation of the matrix $M_i{}^{j(0)}$ 
\equ{matrix-M0} using the 5-bein \equ{5bein-esferico}
(with the non-diagonal component $c(r)=0$)
indeed shows that its rank takes the maximum value, 50,
hence it is regular. We have also computed its 
determinant for the case of the solution \equ{solution-Schw-deS}:
\[
\mbox{Det}\,(M_i{}^{j(0)}) =
-\frac{4608 \mu ^6 \left(2 r^3+{l}^2 \mu \right)^3}
{{l}^{88} r^{27}}\,,
\]
which is clearly not vanishing as long as the mass $\m$ is not
equal to zero.

\section{Cosmological solutions}\label{cosmology-sol}

We turn now to the search for cosmological solutions, 
 again the case 
of a positive cosmological constant $\LA$, \eg taking the parameter 
$s$ equal to 1. 

This search is 
based on
the hypothesis of isotropy and homogeneity  of the physical 3-space.
The space-time coordinates are taken as $t,r,\theta,\f,\chi$ as in 
Section \ref{spherical-sol}, $r,\theta,\f$ being spherical 
coordinates for the 3-space and $\chi\in (0,2\pi)$ the compact subspace $S^1$ coordinate. 
We shall only consider here the zero modes of Kaluza-Klein, \ie all functions will only depend on the time coordinate $t$.

The most general metric
satisfying our symmetry requirements, up to
general coordinate transformations, is given by
Eq. \equ{FLRW-metric-app} of Appendix \ref{app-cosmo}. In the present case of $\chi$-independence, we can perform another time coordinate transformation in order to eliminate the factor in front of $dt^2$,
which yields the metric
\eq
ds^2 =  -dt^2 + \frac{a^2(t)}{1-k r^2} dr^2 + r^2 \lp d\theta^2 
+  \sin^2\theta\, d\f^2 \rp
+b^2(t) d\chi^2\,,
\eqn{FLRW-metric}
which is of the FLRW type in what concerns the 4D sub-space-times 
at constant $\chi$. 
We shall restrict on the case of a flat 3-space, \ie $k=0$. This metric
can then be obtained, using \equ{5D-metric}, from the 5-bein
\eq
\lp e^A{}_\a \rp = \mbox{diag}\lp
1,\,a(t),\,a(t)r,\,a(t)r\sin\theta,\,b(t)\rp \,,
\eqn{FLRW-bein}
 up to a 5D Lorentz transformation.
We shall not assume from the beginning a null torsion $T^A$ \equ{torsion}. 
Due to the isotropy 
and homogeneity conditions, the torsion depends on five 
independent functions $\tilde{f}(t)$, $h(t)$, $\tilde{h}(t)$, $u(t)$ and 
$\tilde{u}(t)$, as shown in Eq. \equ{torsion-app} of 
Appendix \ref{app-cosmo}.
The equations resulting from the field equations \equ{field-eq} are also
displayed in this appendix.

Let us now show that two components of the torsion, namely $u$ 
and $\tilde{u}$,
can be set to zero by a partial gauge fixing condition. The two gauge invariances which are fixed in this way are the ones generated by $P_0$ and
$P_4$, \ie the transformations \equ{gauge-transf} for the parameters
$\b^0(t)$ and $\b^4(t)$. The torsion
components which transform non-trivially are
 $T^0_{t\chi}$ and $T^4_{t\chi}$:
\[
\d T^0_{t\chi} = F^{04}_{t\chi}\b_4\,,\quad
\d T^4_{t\chi} = -F^{04}_{t\chi}\b_0\,.
\]
These transformations are non-trivial as a consequence of the 
non-vanishing of the $F$-curvature component occurring here:
\[
F^{04}_{t\chi} = -\ddot b +\pa_t(\dot{b}\tilde{u})+\frac{b}{l^2}\,,
\]
as it can be read out from  \equ{F-app}.
It follows therefore that the gauge fixing conditions 
\eq
u=0\,,\quad \tilde{u}=0\,,
\eqn{torsion-gauge-fixing}
are permissible. 

We shall describe matter with the perfect fluid energy-momentum
tensor \equ{en-mom-app} with zero pressure, ${p}=0$, a non-vanishing 
energy density and a possibly non-vanishing ``compact dimension pressure'': 
\eq
\TT^A{}_B=\mbox{diag}\,\lp - \dfrac{\rho(t)}{2\pi b(t)},0,0,0,
\dfrac{\la(t)}{2\pi b(t)}\rp\,.
\eqn{en-mom-cosmologia}
The first entry here is the energy density $\hat\rho(t)$
in 4-space, written in terms of the effective 
3-space energy density $\rho(t)$. We have correspondingly redefined 
$\la$ for the sake of homogeneity in the 
notation (see Appendix \ref{not-conv}).
With this form of the energy-momentum tensor and the 
assumptions made at the beginning of this section, the continuity equation 
\equ{FLRWceq2} is trivially satisfied, whereas \equ{FLRWceq1} reads
\eq 
\dot{\rho}
 +3\rho\dfrac{\dot{a}}{a}
 + (\rho+\lambda)\dfrac{\dot{b}}{b} = 0\,, 
 \eqn{FLRWcontinuity}
which  reduces to the usual continuity equation for dust in the case 
of a constant compactification scale $b$.

Let us now solve the field equations (\ref{26} - \ref{35}).
A key observation is that none of both expressions
$(\mathbb{K}-\frac{s}{l^{2}})$ 
and $(\mathbb{B}-\frac{s}{l^{2}})$
can vanish, since we assume a non-zero energy density. (Remember 
the gauge conditions \equ{torsion-gauge-fixing}, and that
all derivatives in $\chi$ vanish since we only consider the 
zero KK modes.) 
Then, Eqs. \equ{35}, \equ{31} and \equ{34}, taken in that order, 
 imply
\[
\tilde{h}(t)=0\,,\quad h(t)=0\,,\quad \tilde{f}(t) = \mbox{constant}\,,
\]
respectively. Now, solving \equ{33} leads to two possibilities: 
$\tilde{f}$ vanishing or not. Let us first show that the latter case
leads to a contradiction.
 \equ{33}  with $\tilde{f}$ $\not=0$  
 implies the equation $\ddot{b}/b-1/l^2$ = 0, which solves 
in $b(t)= b_0 \exp(\pm t/l)$. Then Eq. \equ{30} reads
 $(\ddot{a}-1/l^2)(\dot{a}-1/l)$ which solves in $a(t)= a_0 \exp(\pm t/l)$.
 Inserting this into Eq. \equ{26} yields $\rho(t)=0$,
 which contradicts the hypothesis of a non-vanishing energy density.
We thus conclude that $\tilde{f}(t) =0$,
which finally means a vanishing torsion\footnote{The attentive reader may -- correctly -- find that the gauge fixing conditions 
\equ{torsion-gauge-fixing} are not necessary in order to achieve the result in the case $\tilde{f}=0$: Their are in fact consequences,
together with $h=\tilde{h}=0$,  of the field equations 
(\ref{31}-\ref{35}). But the result above for  $\tilde{f}\not=0$
indeed does need these gauge fixing conditions.}: 
\[
T^A = 0\,.
\]
In order to solve now for the  remaining 
field equations, we make the simplifying hypothesis 
that the compactification scale is constant:
\eq
b(t) = R\,.
\eqn{b=R-2}
At this stage, the field equations reduce to the system
\eq\ba{l}
\dfrac{\dot a(t)^2}{a(t)^2}-\dfrac{\Lambda}{3}
=\dfrac{8 \pi  G}{3} \rho (t)\es
\dfrac{\ddot a(t)}{a(t)}+\dfrac{\dot a(t)^2}{2 a(t)^2}-\dfrac{\Lambda }{2}=0\,,\es
\la(t) = \dfrac{-3 a^2(t) \ddot a(t)
-3 a(t) \dot{a}^2(t)
+\frac{9}{\Lambda} \dot{a}^2(t) \ddot a(t)
+\LA a^3(t)}{16 \pi ^2 G R\, a^3(t)}\,,
\ea\eqn{T=0-b=R-eqs}
where we have expressed the parameters $\kappa$ and $l$ in terms of the Newton constant $G$ and the cosmological constant $\LA$ as
\eq
\kappa = -\dfrac{16\pi^2}{3}G R\LA    \,,\quad
l = \sqrt{\dfrac{3}{\LA}}\,.
\eqn{kappa-G4D}
We recognize in the first two equations the Friedmann equations for dust. The third equation gives the ``compact dimension pressure'' $\la$.

With the Big Bang boundary conditions a(0)=0 ,
the solution of the system reads
\eq\ba{l}
a(t) = C \sinh^{\frac23}\lp \frac{\sqrt{3\LA}}{2}t\rp\,,\quad
\rho(t) = \dfrac{\LA}{8\pi G}\lp\dfrac{C}{a(t)}\rp^3\,,\es
\la(t) = - \dfrac{\LA}{32\pi^2 G R}\lp\dfrac{C}{a(t)}\rp^6\,.
\ea\eqn{cosmol-sol}
where $C$ is an integration constant. The first line of course reproduces
the $\LA$CDM solution for dust matter, whereas the second line
shows a decreasing of $\la$ as the sixth inverse power of the scale parameter $a$.

As it should, the solution obeys the continuity equation
\equ{FLRWceq1}, which now reads
\[
\dot{\rho}+3\rho\dfrac{\dot{a}}{a} = 0\,.
\]
The continuity equation \equ{FLRWceq2} is trivially satisfied.

We recall that we have made the assumption of a constant 
scale parameter $b$
for the  compact dimension. This assumption is not necessary, 
but it is interesting to note, as it can be easily checked, that 
solving the equation in which we insert the 
$\LA$CDM expression of \equ{cosmol-sol} for the 3-space scale parameter $a(t)$,
implies the constancy of $b$.

We have also explicitly checked the validity of the condition 
for stability according to the criterion of Subsection 
\ref{stability-criterion}: the matrix $M_i{}^{j(0)}$ 
\equ{matrix-M0} calculated
 using the 5-bein \equ{FLRW-bein}  has its maximum rank, 50,
 hence it is regular. We have also computed its 
determinant for the case of the solution \equ{cosmol-sol}:
\[
\mbox{Det}\,(M_i{}^{j(0)}) = 
{-\frac{439453125 \left(4-9\, \text{coth}^2
\left(\frac{3 t}{2 {l}}\right)\right)^{16} 
\left(16+9\, \text{coth}^2\left(\frac{3 t}{2
{l}}\right)\right)^3}{1152921504606846976\, l^{100}}}\,,
\]
which is generically not vanishing as a function of $t$.

\section{Conclusions}\label{conclusion}

After recalling basic facts on the five-dimensional Chern-Simons gravity
with the five dimensional (anti)-de Sitter (\ADS) gauge group,
we have studied some important aspects of this theory in comparison with the results of General Relativity with cosmological constant.

First of all, the cosmological constant is here a necessary ingredient
due to the \ADS\ algebraic structure, although it remains a free parameter.
It cannot be set to zero. 

We have shown that, for a spherically symmetrical 3-space, 
the ``vacuum''  Sch\-warz\-schild-de Sitter solution \equ{solution-Schw-deS} 
follows uniquely from the hypotheses of a zero-torsion, 
stationary and static geometry. However, the existence of 
this solution implies the presence of a non-vanishing 
``compact dimension pressure'' 
$\la(r)$ as given by \equ{lambda-spherical}, a fact of non-easy interpretation, in particular due to the expected smallness of the compactification scale.

For the other physically interesting case of a cosmological model 
based on an homogeneous and isotropic 3-space, where we have restricted ourselves to the observationally favored 
flatness of 3-space, we have shown
that the equations for the Friedmann scale parameter $a(t)$ 
and the energy density $\rho(t)$
are identical to the well-known Friedmann equations of General Relativity
under the hypothesis that the compact scale parameter $b(t)$ be a constant. Conversely, only this constancy is compatible with the Friedmann equations.
There is also a non-vanishing ``compact dimension pressure'', decreasing 
in time as the sixth inverse power of the scale parameter $a(t)$. We have also seen that the vanishing of the torsion follows from 
the full \ADS\ gauge invariance and of the field equations.

An important aspect of this work is the establishment of a criterion
guarantying the stability of the zero-torsion solutions if a certain condition based on zero-torsion geometrical quantities is fulfilled. We have also checked that this condition is indeed met in the two situations considered in this paper.

Summarizing all these considerations, we can conclude that
the two families of solutions investigated here coincide with the corresponding solutions of General Relativity in presence of a (positive) cosmological constant. However we recall that we have only examined the Kaluza-Klein zero modes of the theory. 
Possible deviations from the results of Einstein General Relativity  
could  follow from the consideration of higher modes. Also, solutions with torsion would be interesting in view of its possible physical effects.


\subsection*{Acknowledgments}

 Use has been made of the 
 differential geometry computation program ``matrixEDC for 
Mathematica''~\cite{Bonanos} for various calculations. \\
\indent
This work was partially funded by the
Funda\cao\ de Amparo \`a Pesquisa do Estado de Minas Gerais -- 
FAPEMIG, Brazil (O.P.),
the Conselho Nacional de Desenvolvimento Cient\'{\i}fico e
 Tecnol\'{o}gico -- CNPq, Brazil (I.M., Z.O.  and O.P.)
and the Coordena\cao\ de Aperfei\c coamento de Pessoal de N\ii vel 
Superior -- CAPES, Brazil (I.M. and B.N.).

\appendix
\section*{Appendices}

\section{Notations and conventions}\label{not-conv}

\begin{itemize}
\item Units are such that $c=1$.
\item Indices $\a,\,\b,\,\cdots$ = $0,\,\cdots,\,4$, also
called $t,\,r,\,\theta,\,\f,\,\chi$,  are 5D 
space-time coordinates. 
\item Indices.
$A,\,B,\,\cdots$ = $0,\,\cdots,\,4$ are 5-bein frame indices.
\item 
Indices $A,\,\cdots$, are raised or lowered with the Minkowski metric $(\eta_{AB})$ =
 diag$(-1,1,1,1,1)$. 
 \item Indices $\a,\,\cdots$ may be exchanged with
 indices $A,\,\cdots$ using the 5-bein ${e}^A{}_\a$ or its
 inverse ${e}^\a{}_A$.
\item A hat on a symbol means a 5D quantity, like \eg 
$\hat{\rho}(t,x^1,x^2,x^3,x^4)$ for the energy density in 4-space.
\end{itemize}

\section{Construction of the spin 
connection}\label{app-connection}

We recall here how the spin connection ${\om}$ can be constructed 
 from the 5-bein ${e}$ and the torsion 
 ${T}$~\cite{Toloza-Zanelli}.
 First, given the 5-bein, one constructs the torsion-free connection 
 $\bar{\om}$, solution of the zero torsion equation
 $d{e}^A  + \bar{\om}^A{}_B \wedge  {e}^B =0$.
 The result~\cite{Bertlmann} is 
 \[
 \bar{\om}^A{}_{B\m} = \half\lp \xi_C{}^{AB}+\xi^B{}_C{}^A
 -\xi^{AB}{}_C\rp {e}^C{}_\m\,,
 \]
 with
 \[
 \xi_{AB}{}^C= e^\m{}_A e^\n{}_B (\pa_\m e^C{}_\n - \pa_\n e^C{}_\m)\,.
 \]
One then defines the contorsion 1-form  $\CC^A{}_B$ by the equation
$T^A$ = $\CC^A{}_B \wedge {e}^B$, which solves in
\eq
\CC^{AB} = - \half\lp {T}{}^{AB}{}_C+{T}{}_C{}^A{}^B
 -{T}^{B}{}_C{}^A\rp {e}^C{}\,,
\eqn{contorsion}
where ${T}{}^{A}{}_{BC}$ = 
$e^A{}_\m \,T^\m{}_{\n\rho}\, e^\n{}_B\, e^\rho{}_C$
are the torsion components in the 5-bein 
basis\footnote{In the 5-bein basis: ${T}{}^{A} = 
\frac12 {T}{}^{A}{}_{BC}\,{e}^B \wedge {e}^C$.}.
From this we get the full connection form as
\eq
{\om}^A{}_B  =  \bar{\om}^A{}_B + \CC^A{}_B\,,
\eqn{full-connection}
obeying the full torsion equation  $T^A$ =
$d{e}^A  + {\om}^A{}_B \wedge {e}^B =0$.

\section{Metric and torsion for 3-space spherical 
symmetry}\label{app-spher}

In this Appendix, we derive the metric  $g_{\m\n}$ 
and torsion tensors $T^\rho{}_{\m\n}$ in the case of
a 3-space with spherical symmetry around the origin $r=0$.
Accordingly,  observables such as the metric and the torsion components in the coordinate basis must satisfy
Killing equations, which are the vanishing of the Lie derivatives of the fields along the vectors $\xi$ which generate the symmetries.

The set o Killing vectors $\xi$ are the generators $J_{i}$
($i=1,\,2,\,3$) of $SO(3)$, 
which generate the spatial rotations. In the coordinate 
system $t,r,\,\theta,\,\phi,\,\chi$, where $r,\,\theta,\,\phi$
are spherical coordinates for 3-space, and $\chi$ the compact subspace coordinate, these vectors read
\begin{eqnarray*}
J_{1}&=&-\sin{\phi}\partial_{\theta
}-\cot{\theta}\cos{\phi}\partial_{\phi}\,,\\
J_{2}&=&\cos{\phi}\partial_{\theta}
-\cot{\theta}\sin{\phi}\partial_{\phi}\,,\\
J_{3}&=&\partial_{\phi}\,,
\end{eqnarray*}
and obey the commutation rules
\[
\left[J_{i}, J_{j}\right]=\e_{ijk}J_{k}\,,
\]
The Killing equations for the metric and the torsion read,
for $\xi$ = $J_1,\,J_2,\,J_3$,
\begin{eqnarray*}
\pounds_{\xi}g_{\mu\nu}&=&{\xi}^{\rho}\partial_{\rho}g_{\mu\nu}+g_{\rho\mu}\partial_{\nu}{\xi}^{\rho}+g_{\nu\rho}\partial_{\mu}{\xi}^{\rho}=0\,,\\
\pounds_{\xi}T^{\delta}\!_{\mu\nu}&=&{\xi}^{\rho}\partial_{\rho}T^{\delta}\!_{\mu\nu}-T^{\rho}\!_{\mu\nu}\partial_{\nu}{\xi}^{\delta}+T^{\delta}\!_{\rho\nu}\partial_{\mu}{\xi}^{\rho}+T^{\delta}\!_{\mu\rho}\partial_{\nu}{\xi}^{\rho}=0\,.
\end{eqnarray*}
This yields, for the metric:
\eq\ba{l}
d{s}^{2}={g}_{tt}(t,r,\chi) dt^{2}+2{g}_{tr}(t,r,\chi)dt dr +2{g}_{t\chi}(t,r,\chi)dt d\chi+2{g}_{r\chi}(t,r,\chi)dr d\chi \\
\phantom{d{s}^{2}=} +{g}_{rr}(t,r,\chi) dr^{2}+{g}_{\chi\chi}(t,r,\chi) d\chi^{2}+{g}_{\theta\theta}(t,r,\chi)\left( d\theta^{2}+\sin^{2}(\theta)d\phi^{2}\right).
\ea\eqn{metric-t-r-chi}
If we perform a change of radial coordinate 
$r$ to $r'$ = $( {g}_{\theta\theta}(t,r,\chi))^{1/2}$, 
and after that drop the primes, the line element becomes
\begin{eqnarray}
\label{mss}
d{s}^{2}&=&{g}_{tt}(t,r,\chi) dt^{2}
+2{g}_{tr}(t,r,\chi)dt dr +2{g}_{t\chi}(t,r,\chi)dt d\chi
+2{g}_{r\chi}(t,r,\chi)dr d\chi \nonumber\\
\hspace{30pt} && +{g}_{rr}(t,r,\chi) dr^{2}
+{g}_{\chi\chi}(t,r,\chi) d\chi^{2}
+r^{2}\left(d\theta^{2}
+\sin^{2}(\theta)d\phi^{2}\right).
\end{eqnarray}
We shall consider the stationary case, \ie where 
the components of the metric are independent of 
the coordinate  $t$.
For this case we can consider the differential 
${g}_{\chi\chi}(r,\chi) d\chi+{g}_{r\chi}(r,\chi)dr$, 
and from the theory of partial differential equations 
we know that we can  multiply it by an integrating factor 
$I_1=I_1(r, \chi)$  which makes it an exact differential. 
Using this result to define a new coordinate $\chi'$ 
by requiring 
$d\chi'=I_{1}(r,\chi)({g}_{\chi\chi}(r,\chi) d\chi
+{g}_{r\chi}(r,\chi)dr)$, 
substituting this in the latter expression of the 
line element and again dropping the prime, the line 
element simplifies to\footnote{This well known argument may be found  
in the textbook~\cite{dinverno}.},
\begin{eqnarray}
\label{mss0}
d{s}^{2}&=&{g}_{tt}(r,\chi) dt^{2} 
+2{g}_{tr}(r,\chi)dt dr
+2{g}_{t\chi}(r, \chi)dt d\chi \nonumber\\
\hspace{30pt} && +{g}_{rr}(r,\chi) dr^{2}
+{g}_{\chi\chi}(r,\chi) d\chi^{2}
+r^{2}\left( d\theta^{2}+\sin^{2}(\theta)d\phi^{2}\right).
\end{eqnarray}
Now we go to the special case where the components of 
the metric depend only on the radial variable $r$,
which amounts to restrict to the Kaluza-Klein zero modes. 
We  can now consider the differential form 
${g}_{tt}(r) d t +{g}_{tr}(r)dr$ and multiply it by an 
integral factor $I_{2}(t,r)$ that permits to write it 
as a perfect differential,  
$dt'=I_{2}(t,r)({g}_{tt}(r) d t +{g}_{tr}(r)dr)$.
Substituting in the line element and dropping 
the prime we finally get
\eq
d{s}^{2}={g}_{tt}(r) dt^{2} +2{g}_{t\chi}(r)dt d\chi  
+{g}_{rr}(r) dr^{2}+{g}_{\chi\chi}(r) d\chi^{2}
+r^{2}\left( d\theta^{2}+\sin^{2}(\theta)d\phi^{2}\right).
\eqn{mss1}
For the torsion, the Killing equations leave the following 
non-vanishing components:
\eq\ba{lll}
{T}^{t}_{tr}={h}_{1}(t,r,\chi), 
&\quad {T}^{t}_{t\chi}={q}_{1}(t,r,\chi), 
&\quad {T}^{t}_{r\chi}={q}_{2}(t,r,\chi), \\
{T}^{r}_{tr}={h}_{2}(t,r,\chi), 
&\quad {T}^{r}_{t\chi}={q}_{3}(t,r,\chi), 
&\quad {T}^{r}_{r\chi}={q}_{4}(t,r,\chi),\\
{T}^{\chi}_{tr}={h}_{5}(t,r,\chi), 
&\quad {T}^{\chi}_{t\chi}={q}_{5}(t,r,\chi), 
&\quad {T}^{\chi}_{r\chi}={q}_{6}(t,r,\chi),\\
{T}^{t}_{\theta\phi}=\sin(\theta){f}_1(t,r,\chi,
&\quad  {T}^{r}_{\theta\phi}=\sin(\theta){f}_2(t,r,\chi),
&\quad {T}^{\chi}_{\theta\phi}=\sin(\theta){f}_5(t,r,\chi),
\\
{T}^{\theta}_{t\theta}={h}_{3}(t,r,\chi)={T}^{\phi}_{t\phi}, 
&\quad {T}^{\theta}_{t\phi}=\sin(\theta){f}_3(t,r,\chi), 
&\quad  {T}^{\phi}_{t\theta}=-\dfrac{{f}_3(t,r,\chi)}
{\sin(\theta)},\\
{T}^{\theta}_{r\theta}={h}_{4}(t,r,\chi)={T}^{\phi}_{r\phi},
&\quad {T}^{\theta}_{r\phi}=\sin(\theta){f}_4(t,r,\chi),
&\quad {T}^{\phi}_{r\theta}=-\dfrac{{f}_4(t,r,\chi)}{\sin(\theta)}\\
{T}^{\theta}_{\theta\chi}={h}_{6}(t,r,\chi)={T}^{\phi}_{\phi\chi}, 
&\quad  {T}^{\theta}_{\phi\chi}=\sin(\theta){f}_6(t,r,\chi),
&\quad  {T}^{\phi}_{\theta\chi}=-\dfrac{{f}_6(t,r,\chi)}{\sin(\theta)}.
\ea\eqn{ss}

A derivation of the connection $\om(t,r,\chi)$ from the general metric \equ{mss} and the torsion \equ{ss} following the lines of Appendix \ref{app-connection}, hence of the curvature forms and the field equations, may be found in~\cite{elsewhere}. In
the present work we shall restrict to solutions which are independent of $t$ (stationary) and independent of $\chi$ (Kaluza-Klein zero  modes). The metric \equ{mss1} will be used.

\section{Equations in the case of an isotropic and homogeneous 
3-space}\label{app-cosmo}

 We give here the derivation of the 
general set of field equations  with full dependence on the compact dimension coordinate $\chi$,
in the case
of a 5D space-time with an isotropic and homogeneous 3D subspace. All fields are functions of the time coordinate $t$ and the compact coordinate $\chi$. 3-space coordinates are spherical: $r,\,\theta,\,\f$.

\subsection{Metric, 5-bein, torsion and curvature}
The cosmological principle demands that the 3D spatial section of space-time be isotropic and homogeneous. Therefore the fields involved in the model must be compatible with this assumption. 
Isotropy of space-time means that the same observational 
evidence is available by looking in any direction in the universe,
\ie all the geometric properties of the space remain invariant after a rotation. Homogeneity means that at any random point  the universe looks exactly the same.
These two assumptions are translated in Killing equations, which are the vanishing of the Lie derivatives of the fields along the vectors $\xi$ which generate the symmetries.

The set o Killing vectors $\xi$ are the generators $J_{i}$
($i=1,\,2,\,3$) of $SO(3)$, 
which generate the spatial rotations, and the generators of spatial translations $P_{i}$, satisfying the commutation rules     
\[
\left[J_{i}, J_{j}\right]=\e_{ijk}J_{k}\,\quad
\left[J_{i}, P_{j}\right]=\e_{ijk}P_{k}\,\quad
\left[P_{i}, P_{j}\right]= -k\e_{ijk}J_{k}\,,
\]
where $k$ is the 3-space curvature parameter: $k=0,\,1,\,-1$ for plane, 
closed or open 3-space, respectively. In our coordinate system, these vectors read
\begin{eqnarray*}
J_{1}&=&-\sin{\phi}\partial_{\theta
}-\cot{\theta}\cos{\phi}\partial_{\phi}\,,\\
J_{2}&=&\cos{\phi}\partial_{\theta}
-\cot{\theta}\sin{\phi}\partial_{\phi}\,,\\
J_{3}&=&\partial_{\phi}\,,
\end{eqnarray*}
 and
 \begin{eqnarray*}
 P_{1}&=&\sqrt{1-kr^{2}}\left(\sin\theta\cos\phi\,\partial_{r}+\frac{\cos\theta\cos\phi}{r}\partial_{\theta}-\frac{\sin\phi}{r\sin\theta}\partial_{\phi}\right)\,,\\
 P_{2}&=&\sqrt{1-kr^{2}}\left(\sin\theta\sin\phi\,\partial_{r}+\frac{\cos\theta\sin\phi}{r}\partial_{\theta}+\frac{\cos\phi}{r\sin\theta}\partial_{\phi}\right)\,,\\
 P_{3}&=&\sqrt{1-kr^{2}}\left(\cos\theta\,\partial_{r}-\frac{\sin\theta}{r}\partial_{\theta}\right)\,.
 \end{eqnarray*}
The Killing conditions must hold for \ADS\ gauge invariant tensors.
We are interested here  in these conditions for the metric tensor
$g_{\a\b}$ = $\eta_{AB}e^A{}_\a e^B{}_\b$
and the torsion tensor
$T^\g{}_{\a\b}$ = $e^\g{}_A T^A{}_{\a\b}$:
\begin{eqnarray*}
\pounds_{\xi}g_{\mu\nu}&=&
{\xi}^{\gamma}\partial_{\gamma}g_{\mu\nu}
+g_{\gamma\mu}\partial_{\nu}{\xi}^{\gamma}
+g_{\nu\gamma}\partial_{\mu}{\xi}^{\gamma}=0\,,\\
\pounds_{\xi}T^{\delta}\!_{\mu\nu}
&=&{\xi}^{\gamma}\partial_{\gamma}T^{\delta}\!_{\mu\nu}
-T^{\gamma}\!_{\mu\nu}\partial_{\nu}{\xi}^{\delta}
+T^{\delta}\!_{\gamma\nu}\partial_{\mu}{\xi}^{\gamma}
+T^{\delta}\!_{\mu\gamma}\partial_{\nu}{\xi}^{\gamma}=0\,,
\end{eqnarray*}
with $\xi$ = $J_1,\,J_2,\,J_3,\,P_1,\,P_2,\,P_3$.
The Killing conditions for the metric yield the line element
\begin{eqnarray*}
d{s}^{2} &=&{g}_{\alpha\beta}dx^{\alpha}dx^\beta\\
&=&g_{tt}(t,\chi) dt^{2}
+g_{\chi\chi}(t,\chi) d\chi^2  
 +\a(t,\chi)\left(\dfrac{dr^{2}}{1-kr^{2}}+r^{2}d\theta^{2}
+r^{2}\sin^{2}\theta\, d\phi^{2}\right)\\
&& +  2g_{t\chi}(t,\chi)dt d\chi\,.
\end{eqnarray*}
 In the same way as we did in Appendix \ref{app-spher}, we can 
eliminate the cross term in $dt\,d\chi$ through a change of the time
 coordinate defined by~\cite{dinverno}
 \[
 dt' = I(t,\chi)\lp g_{tt}(t,\chi)+g_{t\chi}(t,\chi)\rp \,,
 \]
 where $I(t,\chi)$ is an integrating factor turning the right-hand side  into an 
 exact differential. Dropping the prime and redefining the coefficients we write 
 the resulting line element  as
\eq\ba{l}
d{s}^{2}  = \\
\quad -n^2(t,\chi) dt^{2} + b^2(t,\chi) d\chi^2  
+a^2(t,\chi)\left(\dfrac{dr^{2}}{1-kr^{2}}+r^{2}d\theta^{2}+r^{2}\sin^{2}\theta\, d\phi^{2}\right)\,.
\ea\eqn{FLRW-metric-app}
 The non-vanishing  components of the torsion
left by the Killing conditions are:
\eq\ba{l}
{T}^{t}\!_{t\chi}=u(t,\chi), \hspace{40pt}{T}^{r}\!_{tr}={T}^{\theta}\!_{t\theta}={T}^{\phi}\!_{t\phi}=-h(t,\chi)\es
{T}^{\chi}\!_{t\chi}=\tilde{u}(t,\chi),\hspace{38pt}{T}^{r}\!_{r\chi}={T}^{\theta}\!_{\theta\chi}={T}^{\phi}\!_{\phi\chi}=\tilde{h}(t,\chi)\es
{T}^{r}\!_{\theta\phi}=r^2\sqrt{1-kr^{2}}\sin\theta\tilde{f}(t,\chi), \quad
{T}^{\theta}\!_{r\phi}=-\dfrac{\sin\theta\tilde{f}(t,\chi)}{\sqrt{1-kr^2}}, \es
{T}^{\phi}\!_{r\theta}
=\dfrac{\tilde{f}(t,\chi)}{\sin\theta\sqrt{1-kr^2}}\,.
\ea\eqn{torsion-app}
The 5-bein $e^A{}_\a$ corresponding to the metric\equ{FLRW-metric-app}
may be written in a diagonal form by fixing the 10 local invariances generated by
the Lorentz generators $M_{AB}$ (see Eqs. \equ{AdS-algebra}). The
result is
\eq
\lp{e}^{A}{}_\alpha\rp =
\left(
\begin{array}{ccccc} 
n(t,\chi) & 0 & 0 & 0 & 0 \\
 0 & \dfrac{a(t,\chi)}{\sqrt{1-kr^{2}}} & 0 & 0 & 0 \\
 0 & 0 & a(t,\chi)\,r & 0 & 0 \\
 0 & 0 & 0 &a(t,\chi)r\sin{\theta} &0 \\
0 & 0 & 0 & 0 & b(t,\chi)\end{array}
 \right)\,,
\eqn{5-bein-app}
The 5-bein form $e^A=e^A{}_\a dx^\a$ read
\[\ba{lll}
{e}^{1}=\dfrac{a(t,\chi)}{\sqrt{1-kr^{2}}}dr\,,\quad
&{e}^{2}=a(t,\chi)rd\theta\,,\quad
&{e}^{3}=a(t,\chi)r\sin{\theta}d\varphi,\\[4mm]
{e}^{0}={n}(t,\chi)dt\,,\quad
&{e}^{4} = {b}(t,\chi)d\chi\,.
&
\ea\]
To find the connection compatible with the 5-bein 
\equ{5-bein-app}] and the torsion ${T}^A$ (see \equ{torsion-app}),
\ie a connection ${\om}^{AB}$ such that \equ{torsion} holds,
is a lengthy but well-known procedure 
(see, \eg~\cite{Toloza-Zanelli}), summarized in 
Appendix \ref{app-connection}. 
The result reads, in the 5-bein basis,
\begin{eqnarray*}
{\om}^{04}&=&  
Q{e}^{0}-\tilde{Q}{e}^{4}\,,\quad
{\om}^{0i}= \mathbb{U}{e}^{i}\,,\quad
{\om}^{i4}= \tilde{\mathbb{U}}{e}^{i}\,,\quad(i=1,\,2,\,3)\es
{\om}^{12}&=& -\frac{\sqrt{1-kr^2}}{ar}{e}^{2}
-\frac{\tilde{f}}{2a}\,{e}^{3},\\
{\om}^{13}&=& -\frac{\sqrt{1-kr^2}}{ar}{e}^{3}
+\frac{\tilde{f}}{2a}\,{e}^{2},\\
{\om}^{23}&=& -\frac{\cot\theta}{ar}{e}^{3}
-\frac{\tilde{f}}{2a}\,{e}^{1}\,.
\end{eqnarray*}
where
\eq\ba{l}
Q= \dfrac{u}{b}+ \dfrac{\pa_\chi n}{b\,n} \,,\quad
\tilde{Q} = \dfrac{\tilde{u}}{n} - \dfrac{\pa_t b}{b\,n}\,,\es
\mathbb{U}=\dfrac{1}{n}(H+h)\,,\quad
\tilde{\mathbb{U}}=\dfrac{1}{b}(\tilde{H}+\tilde{h}))\,,\quad
H=\dfrac{\partial_{t}a}{a}\,,\quad
\tilde{H}=\dfrac{\partial_{\chi}a}{a}\,.
\ea\eqn{Q-Qtilde-etc}
From the connection and the 5-bein we can calculate the Riemann 
curvature $R^{AB}$ \equ{Riemann} and the \ADS\ curvature$F^{AB}$ = $R^{AB}$ - 
$\dfrac{s}{l^2} {{e}}^A \wedge {{e}}^B$\equ{FAB}.
The result for the latter reads, in the 5-bein basis,
\begin{eqnarray}
{F}^{04}&=&\left(\dfrac{\mathbb{Q}}{b\,n}-\frac{s}{l^2}\right){e}^{0}{e}^{4}\,,\nonumber\\
{F}^{0i}&=&\left(\mathbb{A}-\frac{s}{l^2}\right){e}^{0}{e}^{i}
-\mathbb{A}_{1}\,{e}^{i}{e}^{4}+\mathbb{U}\frac{\tilde{f}}{2a}\,\e^{i}\!_{jk}\,{e}^{j}{e}^{k}\,,\nonumber\\
{F}^{i4}&=&\left(\mathbb{B}-\frac{s}{l^2}\right){e}^{i}{e}^{4}+ \mathbb{B}_{1}{e}^{0}{e}^{i}+ \tilde{\mathbb{U}}\frac{\tilde{f}}{2a}\,\e^{i}\!_{jk}\,{e}^{j}{e}^{k}\,,\nonumber\\
{F}^{ij}&=&\left(\mathbb{K}-\frac{s}{l^2}\right){e}^{i}{e}^{j}
-\frac{1}{2a n}\partial_t\tilde{f}\,
\e^{ij}\!_{k}\,{e}^{0}{e}^{k}
+\dfrac{1}{2a b}\partial_{\chi}\tilde{f}\,
\e^{ij}\!_{k}\,{e}^{k}{e}^{4}
\,,\label{F-app}
\end{eqnarray}
with
\[\ba{ll}
\mathbb{Q}=-\partial_{\chi}({n}Q)
-\partial_{t}({b}\tilde{Q})\,,\quad
&\mathbb{A}=
\dfrac{1}{a\,n}\partial_{t}\left(a\mathbb{U}\right)
- Q\tilde{\mathbb{U}}\,,\nonumber\es 
\mathbb{A}_{1}=\dfrac{1}{a\,b}\partial_{\chi}
\left(a\mathbb{U}\right)+\tilde{Q}\tilde{\mathbb{U}}\,,\quad
&\mathbb{B}=- \dfrac{1}{a\,n}
\partial_{t}(a\tilde{\mathbb{U}})-\tilde{Q}\mathbb{U}\,,\nonumber\es 
\mathbb{B}_{1}=\dfrac{1}{a\,n}
\partial_{t}(a\tilde{\mathbb{U}})-Q\mathbb{U}\,,\quad
&\mathbb{K}=\dfrac{k}{a^2}+\mathbb{U}^2
-\tilde{\mathbb{U}}^2-\left(\dfrac{\tilde{f}}{2a}\right)^2\,.\nonumber
\ea\]
We will also need the torsion components ${T}^A{}_{BC}$
= ${e}^A{}_\a{e}^\b{}_B{e}^\g{}_C{T}^\a{}_{\b\g}$
in the 5-bein basis:
\[\ba{l}
{T}^0{}_{04} = \dfrac{u}{b} \,,\quad
{T}^{4}_{04} = \dfrac{\tilde{u}}{n}\,,\es
{T}^i{}_{0i} =\dfrac{h}{n}\,,\quad
{T}^i{}_{i4} =\dfrac{\tilde{h}}{b}\,,\quad
{T}^i{}_{jk}  = \e^{i}\!_{jk}\,\dfrac{\tilde{f}}{2a}\,,
\quad(i,\,j,\,\cdots,\,=1,\,2,\,3)\,.
\ea\]
As a -2-form, the torsion reads
\begin{eqnarray}
{T}^{0}&=&\dfrac{u}{b} {e}^{0}\wedge{e}^{4}\,,\nonumber\\
{T}^{4}&=& \dfrac{\tilde{u}}{n} {e}^{0}\wedge{e}^{4}\,,\\
{T}^{i}&=&
\dfrac{h}{n}{e}^{0}\wedge{e}^{i}
+\dfrac{\tilde{h}}{b}{e}^{i}\wedge{e}^{4}
+ \dfrac{\tilde{f}}{2a}{e}^{j}\e^{i}\!_{jk}\,\wedge{e}^{k}\,.\nonumber
\end{eqnarray}

\subsection{Field equations}

Matter will be assumed to consist in a spinless perfect fluid 
described by the energy-momentum 4-form
\[
{\TT}_A =\frac{1}{4!}\e_{BCDEF} {\TT}^B{}_A \,
{{e}}^C \wedge {{e}}^D \wedge {{e}}^E \wedge {{e}}^F\,,
\]
with\footnote{ The hats on $\hat\rho$,  etc., mean energy density, etc in 4-space.} 
\eq
\TT^A{}_B=\mbox{diag}\,
(-\hat \rho(t,\chi),\hat p(t,\chi),\hat p(t,\chi),\hat p(t,\chi),
\hat \la(t,\chi))\,,
\eqn{en-mom-app}
and the spin 4-form ${\SSS}_{AB}=0$.

With the expressions above for the curvature and torsion components and
for the matter content,
we can now write the explicit form of the field equations 
\equ{field-eq}:
\begin{eqnarray}
	\label{26}
&& \left(\mathbb{K}-\frac{s}{l^{2}}\right)
\left(\mathbb{B}-\frac{s}{l^{2}}\right)
+\dfrac{1}{2a^2b}\tilde{\mathbb{U}}\tilde{f}
\partial_{\chi}\tilde{f}-\dfrac{l}{24\kappa}\hat{\rho}=0\,,\\
	\label{27}
&&\left(\mathbb{K}-\frac{s}{l^{2}}\right)\mathbb{B}_{1}
-\dfrac{1}{2a^2n}\tilde{\mathbb{U}}\tilde{f}
\partial_{t}\tilde{f} = 0\,,\\
	\label{28}
&& \left(\mathbb{K}-\frac{s}{l^{2}}\right)
\left(\mathbb{A}-\frac{s}{l^{2}}\right)
-\dfrac{1}{2a^2n}\mathbb{U}
\tilde{f}\partial_{t}\tilde{f}+\dfrac{l}{24\kappa}\hat{\lambda} = 0\,,\\
	\label{29}
&& \left(\mathbb{K}-\frac{s}{l^{2}}\right)\mathbb{A}_{1}
-\dfrac{1}{2a^2b}\mathbb{U}\tilde{f}
\partial_{\chi}\tilde{f} = 0\,,\\
   	\label{30}
&&\left(\mathbb{K}-\frac{s}{l^{2}}\right)
\left(
\dfrac{1}{b\,n}\mathbb{Q}-\frac{s}{l^2}\right)
+2\left(\mathbb{A}-\frac{s}{l^{2}}\right)
\left(\mathbb{B}-\frac{s}{l^{2}}\right)
+2 \mathbb{A}_{1}\mathbb{B}_{1} 
   +\dfrac{l}{8\kappa}\hat{p} = 0\,,\quad\quad\\
	\label{31}
&& \left(\mathbb{K}-\frac{s}{l^{2}}\right)
\dfrac{\tilde{u}}{n}
+2\left(\mathbb{B}-\frac{s}{l^{2}}\right)
\dfrac{h}{n} 
-2 \mathbb{B}_{1}\dfrac{\tilde{h}}{b} = 0\,,
\\
	\label{32}
&& \left(\mathbb{K}-\frac{s}{l^{2}}\right)
\dfrac{u}{b} 
+2\left(\mathbb{A}-\frac{s}{l^{2}}\right)
\dfrac{\tilde{h}}{b}+2\mathbb{A}_{1}\dfrac{h}{n} = 0\,, \\
	\label{33}
&& \frac{\tilde{f}}{a}
\left(\dfrac{u}{b} \tilde{\mathbb{U}}
+\dfrac{\tilde{u}}{n}\mathbb{U} 
-\dfrac{1}{b\,n}\mathbb{Q}
+\frac{s}{l^{2}} \right) = 0\,,\\
	\label{34}
&&\left(\mathbb{K}-\frac{s}{l^{2}}\right)
\dfrac{h}{n}
-\dfrac{1}{2a^2n}\tilde{f}\partial_{t}\tilde{f} = 0\,,\\
	\label{35}
&&\left(\mathbb{K}-\frac{s}{l^{2}}\right)
\dfrac{\tilde{h}}{b}
+\dfrac{1}{2a^2b}\tilde{f}\partial_{\chi}\tilde{f} = 0\,.
\end{eqnarray}

\subsection{Continuity equations}
\label{FLRWceq}

For the spinless perfect fluid considered in the latter Subsection, the 
continuity equation \equ{spinless-cont-eq}, consequence of the field equations, takes the form of a system of two equations:
\eq
\partial_{t}\hat{\rho}
 +3 \lp \hat{\rho}+\hat{p}\rp\dfrac{\partial_{t} {a}}{a}
 +\lp \hat{\rho}+\hat{\lambda}\rp\dfrac{\partial_{t} {b}}{b}
 + 3\hat{p}h  - \hat{\lambda}\tilde{u} = 0\,, 
 \eqn{FLRWceq1}
\eq
\partial_{\chi}\hat{\lambda}+3\lp \hat{\lambda}-\hat{p}\rp \dfrac{\partial_{\chi}a}{a}
 + \lp \hat{\rho} + \hat{\lambda} \rp \dfrac{ \partial_{\chi}{n}}{n}
 -  3\hat{p}\tilde{h} + \hat{\rho} u = 0\,.
\eqn{FLRWceq2}


\end{document}